\documentclass[]{spie}  %>>> use for US letter paper
 \usepackage[detect-all]{siunitx}
\usepackage{amsmath,amsfonts,amssymb}
\usepackage{graphicx}
\usepackage[colorlinks=true, allcolors=blue]{hyperref}

\title{Laboratory characterization of iLocater}

\author[a]{Marcelo Tala Pinto}
\author[a]{Marshall C. Johnson}
\author[a]{Julia Brady}
\author[a, g]{Xavier Lesley-Saldaña}
\author[a]{Jonathan Crass}
\author[a]{Allie Renshaw}
\author[a]{Michael Engelman}
\author[b]{Brian Sands}
\author[a]{Daniel Pappalardo}
\author[c]{Christian Schwab}
\author[d]{Julian Stürmer}
\author[c]{Dane Zielinski-Nicolson}
\author[e]{Stanimir Letchev}
\author[d, h]{Matheus J. Castro}
\author[f]{Thomas Legero}
\author[c]{Ondrej Kitzler}
\author[e]{Jacob Pember}
\author[a, g]{Richard Pogge}
\author[b]{Justin R. Crepp}
\author[b]{Andrew Bechter}
\author[b]{Eric Bechter}

\affil[a]{Department of Astronomy, The Ohio State University, 140 W 18th Ave, Columbus, OH, USA}
\affil[b]{University of Notre Dame, Department of Physics and Astronomy, Notre Dame, IN, USA}
\affil[c]{School of Mathematical and Physical Sciences, Macquarie University, Balaclava Road, North Ryde, NSW 2109, Australia}
\affil[d]{Landessternwarte, Zentrum für Astronomie der Universität Heidelberg, Königstuhl 12, 69117, Heidelberg, Germany}
\affil[e]{Max Planck Institute for Astronomy, Königstuhl 17, 69117, Heidelberg, Germany}
\affil[f]{Physikalisch-Technische Bundesanstalt, Bundesallee 100, 38116, Braunschweig, Germany}
\affil[g]{Center for Cosmology and AstroParticle Physics, The Ohio State University, 191 West Woodruff Avenue, Columbus, OH 43210}
\affil[h]{Fakultät für Physik und Astronomie, Universität Heidelberg, Im Neuenheimer Feld 226, 69120 Heidelberg, Germany}

\authorinfo{Further author information: (Send correspondence to M.T.P.)\\M.T.P.: E-mail: tala.1@osu.edu}

\begin{document} 
\maketitle

\begin{abstract}
iLocater is a diffraction-limited, fiber-fed spectrograph designed for the Large Binocular Telescope (LBT), which targets high-precision radial velocity measurements in the near-infrared. Prior to deployment, comprehensive laboratory characterization was essential to validate instrument performance and inform alignment strategies. This paper presents results from four key areas of lab characterization: (1) adjustment and optimization of detector orientation to optimize spectrum alignment with the detector pixel grid across the focal plane; (2) the design and installation of a LED illumination source to enable high-fidelity flat-fields; (3) a model-based focusing methodology using OpticStudio image simulations to optimize the optical alignment of the spectrograph; and (4) assessment of instrument mechanical and optical stability under laboratory conditions. Together, these efforts established baseline performance metrics and demonstrated instrument readiness for delivery and on-sky commissioning.
\end{abstract}


\section{INTRODUCTION}
\label{sec:intro}  % \label{} allows reference to this section

iLocater is a next-generation cryogenic near-infrared (NIR) spectrograph being developed for the Large Binocular Telescope, designed to achieve extreme-precision radial velocity (EPRV) measurements of low-mass stars and potentially detect Earth-like exoplanets\cite{Crepp2016, Crass2022}. Operating in a diffraction-limited, single-mode fiber-fed regime, iLocater represents a significant technical change over conventional seeing-limited spectrographs\cite{Crepp2014}. Before the instrument was deployed at the telescope, thorough laboratory characterization at The Ohio State University (OSU) was required to verify optical performance, establish calibration procedures, and assess long-term stability. This paper describes the lab characterization and optimization campaign conducted at OSU over a series of instrument cooldowns. These focused on four interconnected activities: detector tilt optimization, flat-field illumination system installation, model-based optical focusing, and instrument stability analysis. The results presented here informed the final alignment and calibration strategies ahead of on-sky deployment.

\section{INSTRUMENT OVERVIEW}

iLocater is designed to operate across the Y and J near-infrared bands, fed by single-mode fibers at the focal plane of the LBT's adaptive optics system\cite{Esposito2011, Crass2021}. The spectrograph employs an R6 echelle grating in a cross-dispersed configuration, producing high spectral resolution (R $\sim$ 200,000) spectra from three input fibers that is suitable for detailed stellar characterization and precision RV work. The instrument is housed in a stabilized vacuum cryostat to minimize thermal noise contributions from the surrounding environment while cooling the instrument to its operating temperature of \numrange[range-phrase = --]{80}{100}K\cite{Crass2022_cryo}. Laboratory characterization at OSU employed a dedicated test bench equipped with a fiber injection system, calibration light sources, and environmental monitoring hardware, allowing systematic evaluation of optical and mechanical performance prior to telescope delivery and integration.

\section{DETECTOR CLOCKING}
\begin{figure}
    \centering
    \includegraphics[width=0.95\linewidth]{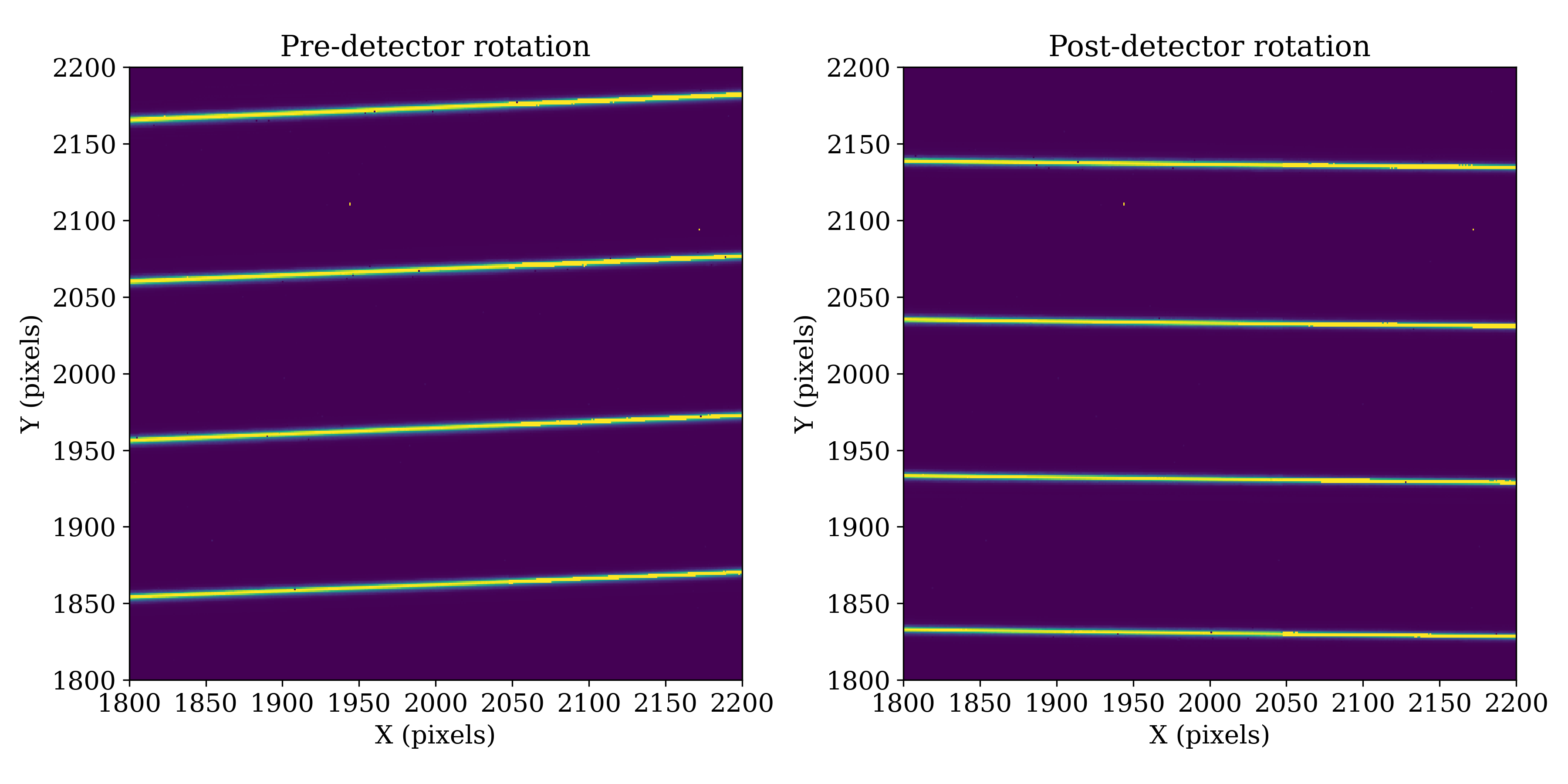}
    \caption{\textit{Left}: Spectral order traces before rotating the detector. \textit{Right}: spectral order traces after rotating the detector.}
    \label{fig:detrot}
\end{figure}

As predicted by iLocater's optical design, the instrument spectrum is tilted with respect to the vertical axis of the spectrograph system. To optimize the optical quality of the spectra on the pixel grid of the detector, we initially assessed spectra orientation during an initial instrument cooldown, and then rotated the detector by a small amount using existing detector mount adjustment features. This improved the alignment of the detector grid with respect to the spectral orders. To quantify how much we needed to rotate the detector, we fitted lines to each of the spectral orders, and calculated the orientation angles with respect to the horizontal axis using the slopes of the linear fits. The angle of rotation adjustment is the mean of the orientation angles averaged across all spectral orders, which corresponds to about 2$^{\circ}$. Figure \ref{fig:detrot} shows a portion of a Halogen spectrum before and after we rotated the detector.

\section{FLAT FIELD ILLUMINATION}
Flat-field calibration is essential for correcting pixel-to-pixel response non-uniformities and intra-order illumination variations. For a cryogenic, fiber-fed instrument like iLocater, integrating an appropriate flat-field illumination source requires careful engineering to ensure uniform, repeatable illumination without introducing stray light or thermal loading. The installed system for iLocater provides sufficient illumination for high signal-to-noise flat frames within practical exposure times and meets the instrument's stray light and thermal requirements.

We use the Thorlabs LED1050E LED coupled to a Thorlabs DG05-120 ground glass diffuser that acts as a flat illumination source. It is mounted off-axis and pointed directly at the detector. Its position is such that the mechanical mounting of the LED and the diffuser do not interfere with the spectrograph's beam, while going through the light baffles placed before the detector illuminating it completely. 

\begin{figure}
    \centering
    \includegraphics[width=0.41\linewidth]{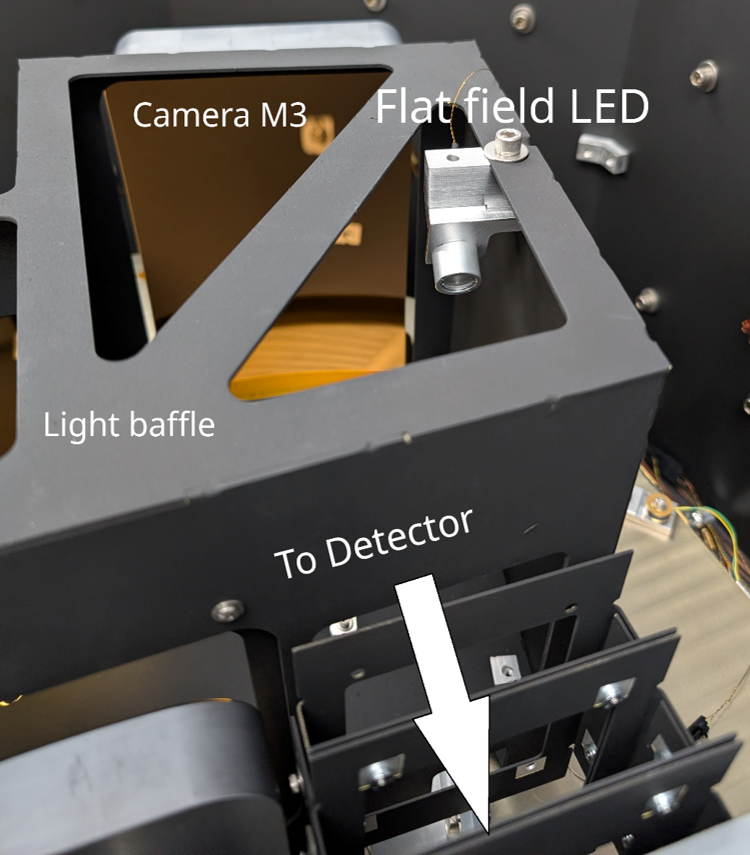}\hspace{0.5cm}\includegraphics[width=0.45\linewidth]{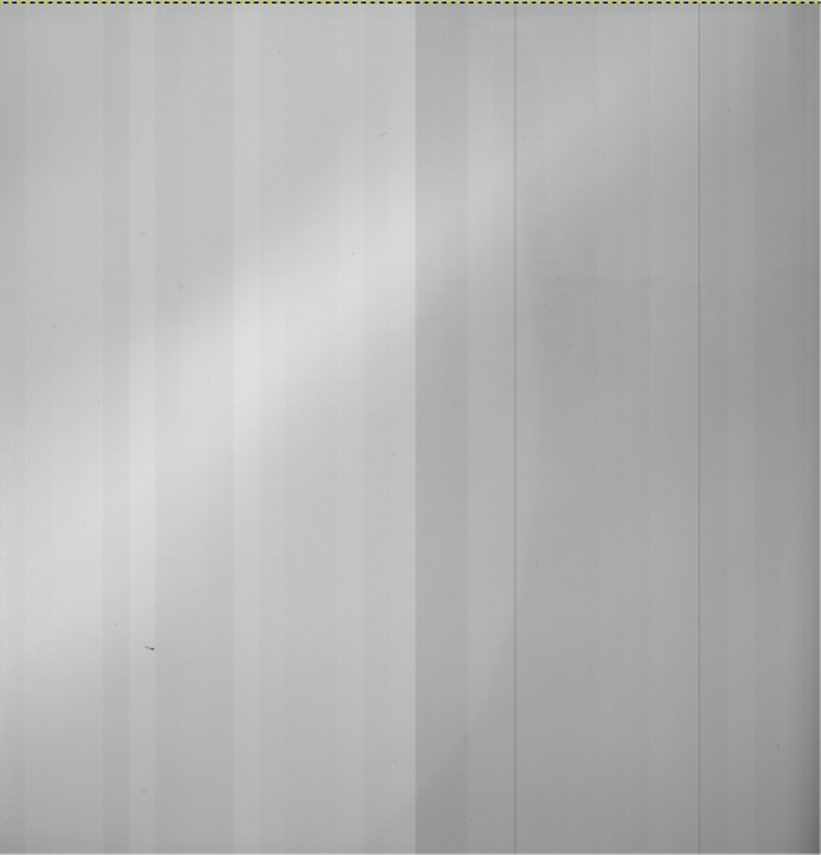}
    \caption{\textit{Left}: picture of the flat field LED mounted on the light baffle pointing towards the detector. \textit{Right}: flat field image obtained with the iLocater detector.}
    \label{fig:led}
\end{figure}

The left panel in Figure \ref{fig:led} show the flat field LED mounted on the top surface of the light baffle for the camera M3. The right panel in Figure \ref{fig:led} show an example of a flat field image obtained with iLocater detector.

\section{FOCUS OPTIMIZATION}
\label{sec:sections}

Achieving optimal focus across the detector is critical for maximizing spectral resolution and minimizing cross-talk between adjacent traces and echelle orders. Rather than relying solely on empirical through-focus measurements — which can be time-intensive for a cryogenic instruments due to thermal cycles being needed for adjustment — we developed a model-based approach using Zemax OpticStudio to guide the focusing procedure. Simulated point spread functions (PSFs) and image spot diagrams were generated as a function of focus position and compared directly to images acquired with the detector. 

\subsection{Image simulations with OpticStudio}
\begin{figure}
    \centering
    \includegraphics[width=0.55\linewidth]{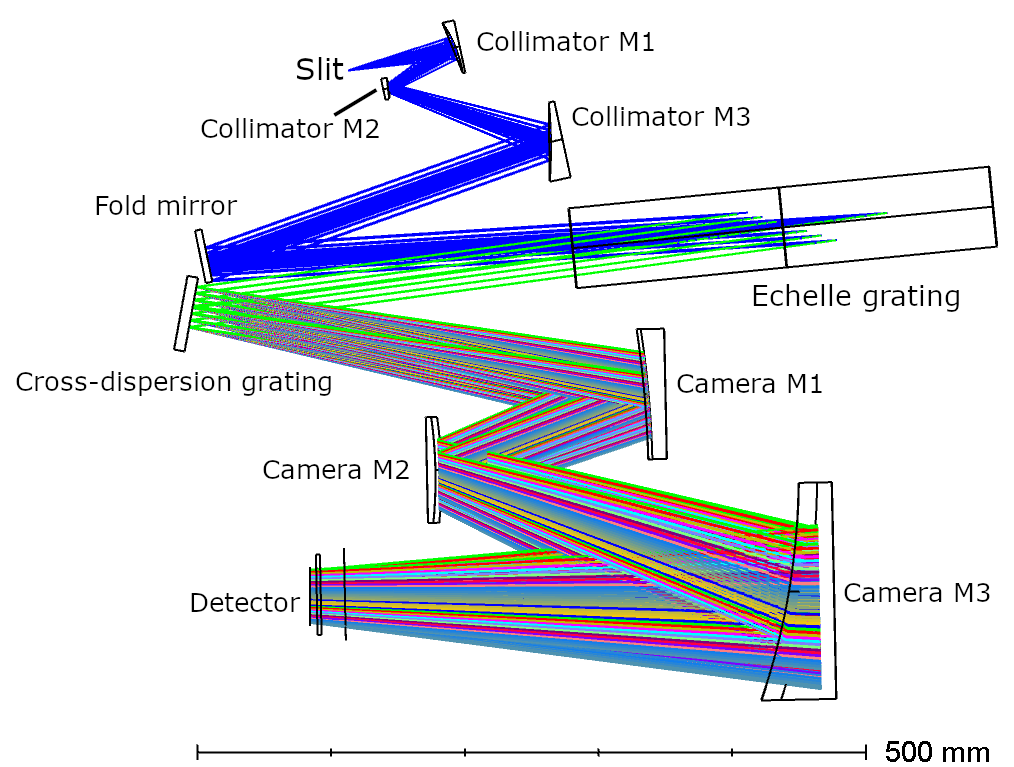}\includegraphics[width=0.45\linewidth]{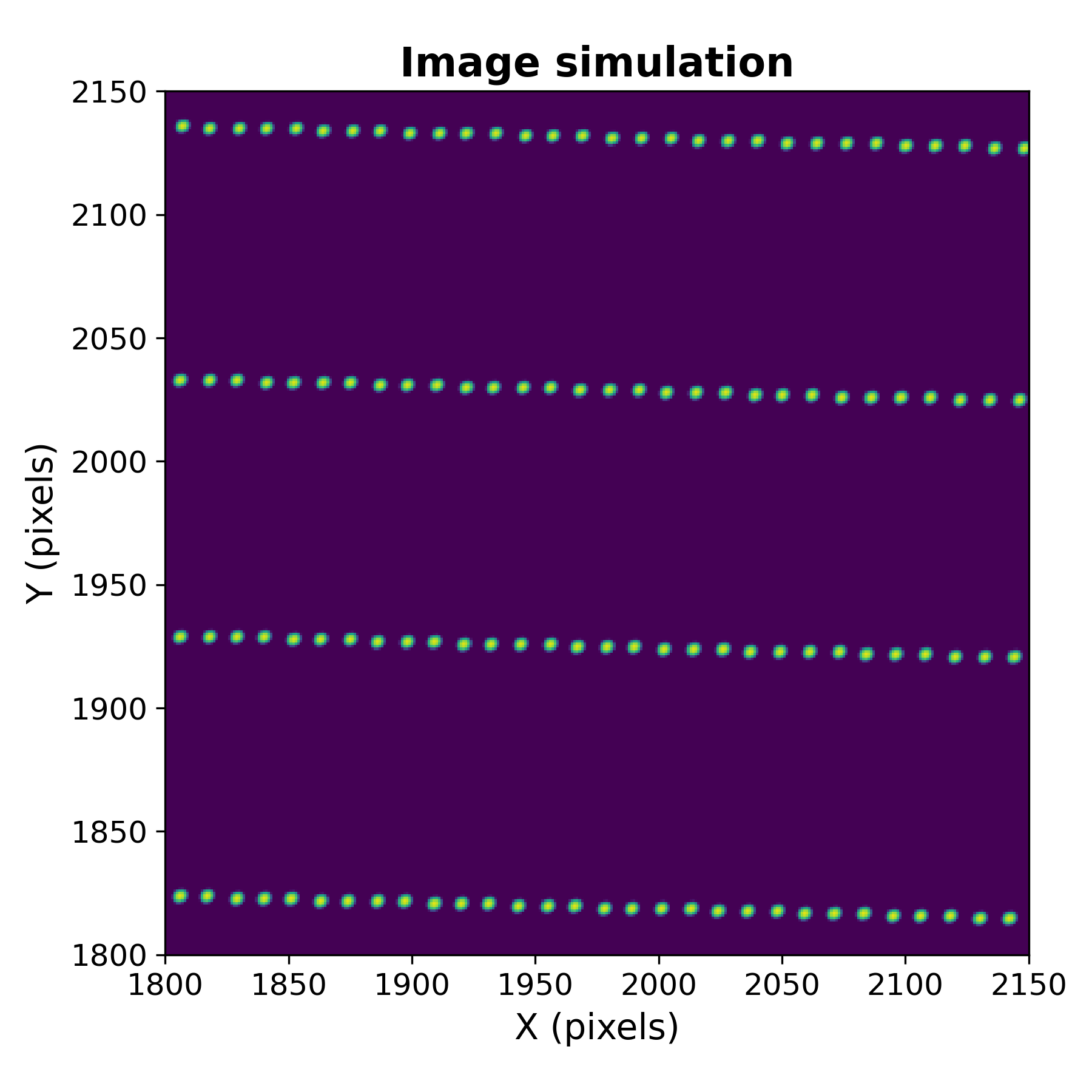}
    \caption{\textit{Left}: iLocater optical layout. Rays after the reflection on the echelle grating and the cross-disperser are color-coded by spectral order. \textit{Right}: spectrum simulated using OpticStudio and \texttt{zospy}. We observe a series of wavelengths traced over four spectral orders located in the central section of the detector.}
    \label{fig:ilocater_optics}
\end{figure}

The left panel in Figure \ref{fig:ilocater_optics} shows the optical layout of the spectrograph. Optically, the spectrograph consists of: (1) an entrance slit with three fibers; (2) a collimator consisting of three aspheric mirrors; (3) a fold mirror to re-direct the collimated beam towards the echelle grating; (4) an echelle grating; (5) a cross-dispersion grating; (6) a camera consisting of three spheric mirrors; (7) a transmissive bandpass filter; and (8) a 4k$\times$4k detector. 

The image modeling was performed using OpticStudio through the Python interface \texttt{zospy}\cite{zospy24}. A two-dimensional grid in spectral order $m$ and wavelength $\lambda$ was sampled across the detector, and for each (m, $\lambda$) coordinate the corresponding image centroid was extracted from the optical model. PSFs were then computed at each sampled field point using OpticStudio and positioned on the detector according to their associated centroids. By combining the centroid locations with the locally evaluated PSFs, a synthetic detector image was generated that captures both the spectral format geometry and the spatial variation of the instrument response across the focal plane. The right panel in Figure \ref{fig:ilocater_optics} shows a central section of a simulated image showing a series of wavelengths traced over four spectral orders.

We simulated images for different focus positions, by adjusting the \textit{thickness} value of the surface before the detector. To incorporate temperature dependent focus effects, we included a model of thermal expansion for the different model parameters, namely: distances between the optical elements, X- and Y-decenters, and curvature radii for curved mirrors. The spectrograph mounts are made of Invar, therefore the distances and decenters are adjusted according to the linear expansion of Invar. The mirrors are made of silicon, therefore the curvature radii are adjusted according to the linear expansion of silicon. We also simulated two different Y-tilts in the detector orientation, to study its effect over the detector focusing.

\begin{figure}
    \centering
    \includegraphics[width=0.95\linewidth]{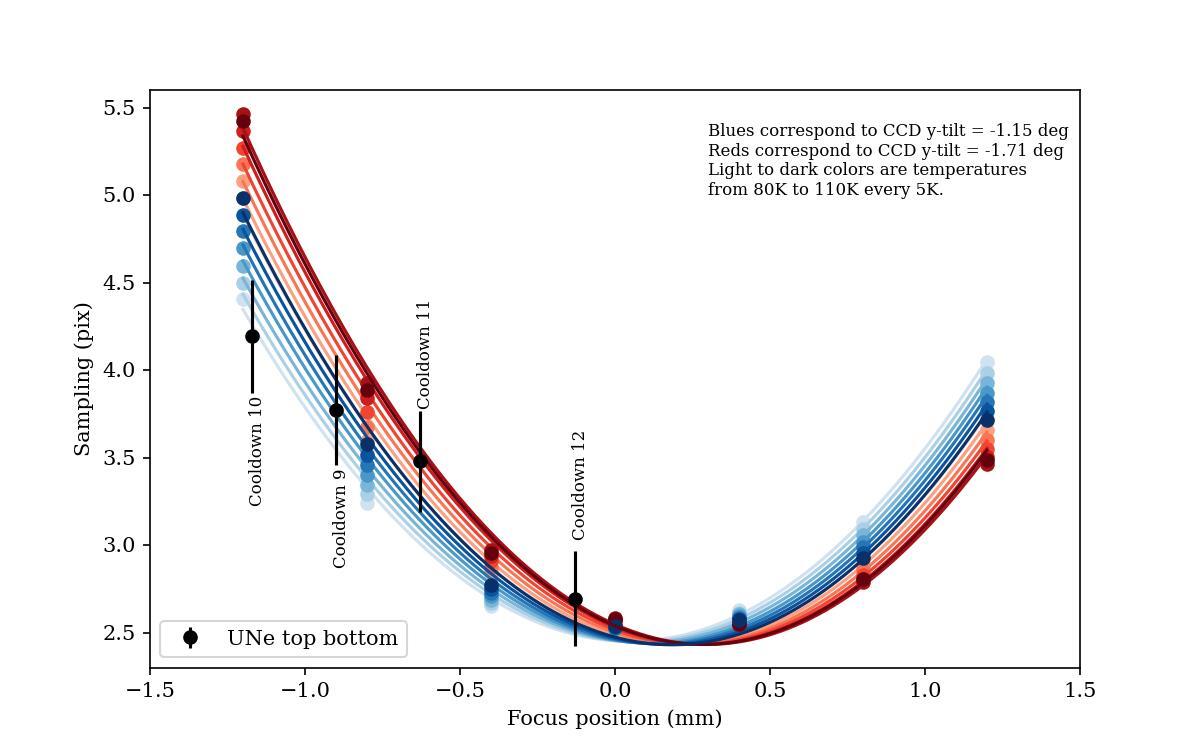}
    \caption{Sampling (in pixels) as a function of focus position (mm) for two detector Y-tilt configurations. Model parabolas are shown in blue (y-tilt = -1.15$^\circ$) and red (y-tilt = -1.71$^\circ$), with color lightness encoding temperature from 80 K (light) to 110 K (dark) in 5 K increments. Black points with error bars indicate measured UNe lamp (top/bottom) sampling from four cooldown epochs (cooldowns 9–12), demonstrating the shift in best-focus position across cooldowns. The minimum sampling of $\sim$2.6 pixels near focus position 0 mm is consistent with the model predictions at intermediate temperatures.}
    \label{fig:focus_parabolas}
\end{figure}

For each model, we simulated an image using a grid of wavelengths (see Figure \ref{fig:ilocater_optics}), and then measured the mean spectral sampling $s$ of the detected peaks, which corresponds to the mean of the FWHM in the X and Y directions of their 2D gaussian fits. We fitted a quadratic relation between focus position and sampling to derive the focus parabolas for each temperature and y-tilt position.
Figure \ref{fig:focus_parabolas} summarizes these results. The two y-tilt models are shown in red and blue, with color lightness encoding temperature from 80 K (light) to 110 K (dark) in 5 K increments.

\subsection{Focusing process}
%The instrument underwent multiple cryogenic thermal cycles during laboratory characterization. Throughout this paper, each cycle is referenced by its sequential cooldown number.

During instrument cooldown 9, an initial sampling of the PSFs across the instrument detector was measured by illuminating the spectrograph with a uranium-neon (UNe) source, however the absolute detector position along the focus axis was not known a priori. 
To assess the focus position of the detector, we identified the centroids of the UNe spectral lines in the image and performed a 2D gaussian fit to the detected lines which was compared to the previously generated model. 
The detector was then stepped in position in successive cooldowns with the system PSF being assessed to allow us to approach best focus: by 270 $\mu$m to the front of the detector from cooldown 9 to 10, by 540 $\mu$m to the back of the detector from cooldown 10 to 11, and by a further 500 $\mu$m to the back of the detector from cooldown 11 to 12. 

To place the measurements in context, the focus position offset was adjusted to match the model parabolas, using the predicted sampling as a function of focus position and temperature as a reference.
While each adjustment moved the system along the steep wing of the parabola, the model predictions at each step guided the magnitude of the correction required in the next cooldown. By cooldown 12, the measured sampling had converged near the minimum of the parabola at approximately -0.1 mm, consistent with the model predictions. 

\subsection{Measured spectral resolution and bandpass}
\begin{figure}
    \centering
    \includegraphics[width=0.9\linewidth]{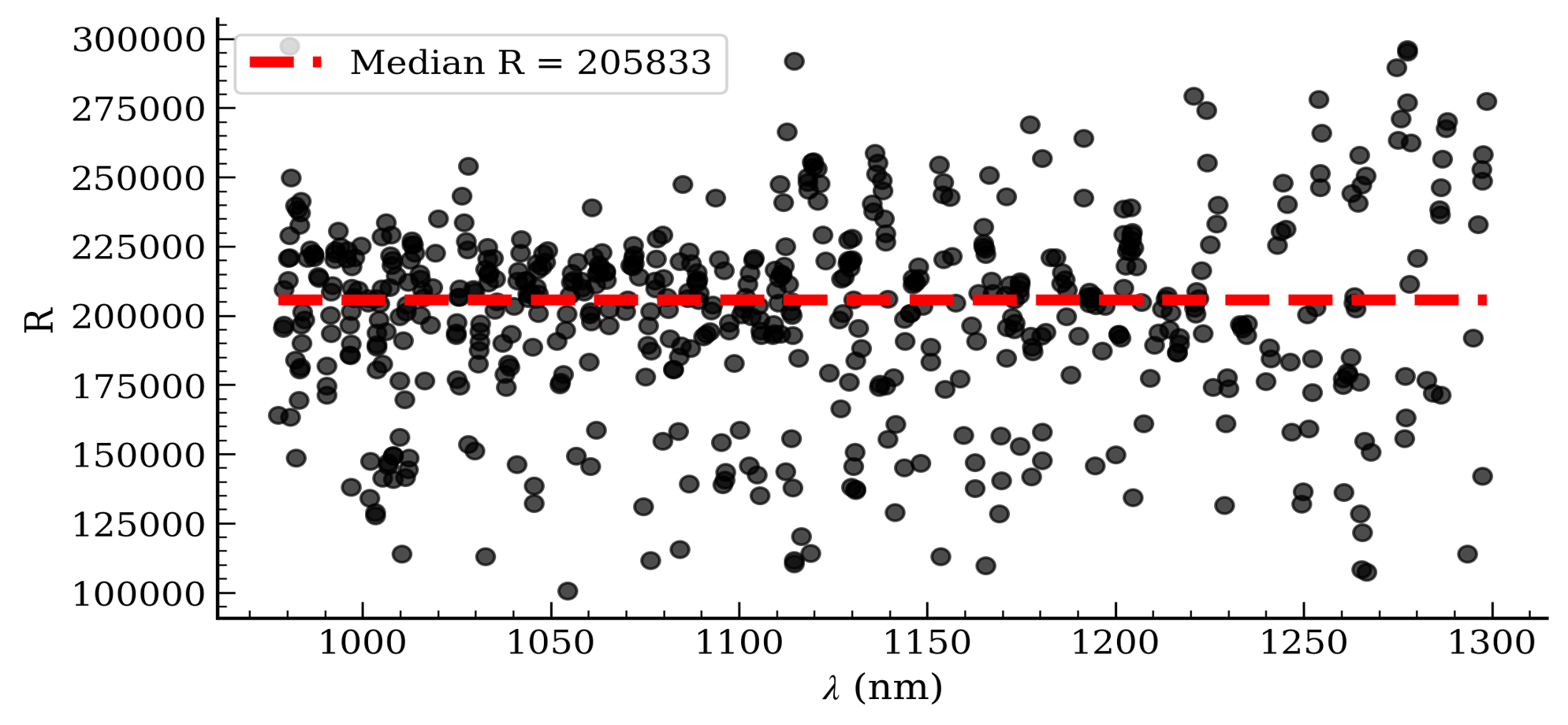}
    \caption{Measured spectral resolution as a function of wavelength.}
    \label{fig:R}
\end{figure}

We measured the spectral resolution and wavelength coverage of the spectrograph by observing the spectra of a UNe lamp. Figure \ref{fig:R} shows the mean spectral resolution for the selected set of UNe lines.

The spectrograph is optimized for the Y-band, and it provides a spectral coverage between 966nm to 1312nm with continuous coverage blueward of 1100nm, across 41 spectral orders, based on the wavelength solution calculated using the procedure described in Section \ref{sec:vel_space} . We measured a mean spectral resolution of R$\sim$205,000 (Figure \ref{fig:R}).

\section{INSTRUMENT THERMAL VS OPTICAL STABILITY}
\subsection{Calibration Data and Reduction}
To characterize instrument stability, we analyzed a series of daily calibration frames acquired each weekday during cooldown 13. Daily calibrations include dark, hollow-cathode uranium–neon (UNe) lamp, unstabilized etalon, and halogen (broadband) frames. Among these, the UNe lamp provided the most stable and reliable wavelength reference and is therefore best suited for assessing RV performance. Although the UNe spectrum contains fewer usable spectral lines than the etalon, it was the most stable source available during testing and is sufficient to demonstrate the RV stability of the instrument.

For this analysis we use UNe calibration data acquired over two weeks, spanning 28 January – 10 February 2026, during which the spectrograph temperature — monitored via a sensor on the echelle grating mount — drifted by approximately 1.5 mK (Figure \ref{fig:rvstability}).

Raw calibration frames are processed from 3-D up-the-ramp datacubes to 2-D images using the HxRGproc package\cite{HxRGproc_Rauscher, HxRGproc_Ninan}. 1-D spectra are then extracted for each of the three input fibers/traces, per spectral order, using a custom implementation of optimal extraction\cite{Horne1986}.

%We use two independent methods to measure spectrograph drift from the extracted 1-D UNe spectra.

\subsection{Individual lines drift}
For each frame, we fit a Gaussian profile to a common set of UNe lines identified across all frames in the analysis period, in each of the three input fibers/traces. 
Lines are retained if their Gaussian fit has a full width at half maximum (FWHM) between 2 and 10 pixels and a peak brightness greater than 5 times the read noise. 
This selection yields between 200 and 700 lines per fiber, with the range driven by differential fiber throughput and the resulting signal-to-noise ratio (SNR) in each fiber.
The line position is taken as the Gaussian centroid, with uncertainty $\sigma$ derived from the fit covariance matrix.

Daily drift is computed as the weighted mean of the per-line positional offsets relative to a reference epoch (28 January 2026), with each line weighted by the inverse variance of its fitted position to account for differences in line strength and measurement precision.

%\subsubsection{Mask-based cross-correlation method}
%Using the UNe spectrum from the reference epoch (28 January 2026), we construct a reference mask for each fiber and order by isolating the core of each selected spectral line — defined as the three pixels on either side of the line peak (Figure 3, left panel) — and setting all other pixels to zero. 
%This approach follows standard cross-correlation mask methodology (e.g., Lafarga et al. 2020). 
%Line selection uses the same FWHM and SNR criteria as above. 
%For each date, we cross-correlate each order's spectral trace against its corresponding reference mask and fit a Gaussian to the resulting cross-correlation function (CCF). 
%The drift for each order/trace is given by the centroid of this Gaussian fit, with uncertainty from the fit covariance matrix. 
%An overall drift and error are then computed as the inverse-variance-weighted mean across all orders for each trace (Figure 3, right panel).

\subsection{Transformation to Velocity Space}\label{sec:vel_space}
The individual drift measurements are in pixel space, however, to assess RV stability, these must be converted to velocity space — a nontrivial step, since the velocity span per pixel varies both within and across orders and therefore requires a wavelength solution. To achieve this, we have used a preliminary wavelength solution, derived by identifying UNe lines of known wavelength\cite{redman2011} and fitting a quadratic dispersion function per order. This solution is accurate to approximately 0.02 nm (10–20 pixels) in absolute wavelength and does not fully capture the dispersion behavior near order edges, particularly given the limitations of a quadratic fit. While insufficient for direct cross-correlation in velocity space, it is adequate for converting individual line-shift measurements to velocity units.

For each line at pixel position $x_i$, the local velocity scale $s_i$ (m/s per pixel) is computed as:
\begin{equation}
    s_i=\frac{\Delta\lambda}{\lambda_i}c
\end{equation}
where $\Delta\lambda=\lambda_i-\ \lambda_{i-1}$ 
is the difference between the wavelengths of adjacent pixels, $\lambda_i$ is the wavelength at pixel i, and c is the speed of light in m/s.  The wavelength solution is interpolated for non-integer pixel positions. The RV drift for a given line is then the pixel-space drift multiplied by $s_i$. Per-fiber velocity drift and its uncertainty are computed as the inverse-variance-weighted mean over all lines.

\subsection{Stability Results}
\begin{figure}
    \centering
    \includegraphics[width=0.9\linewidth]{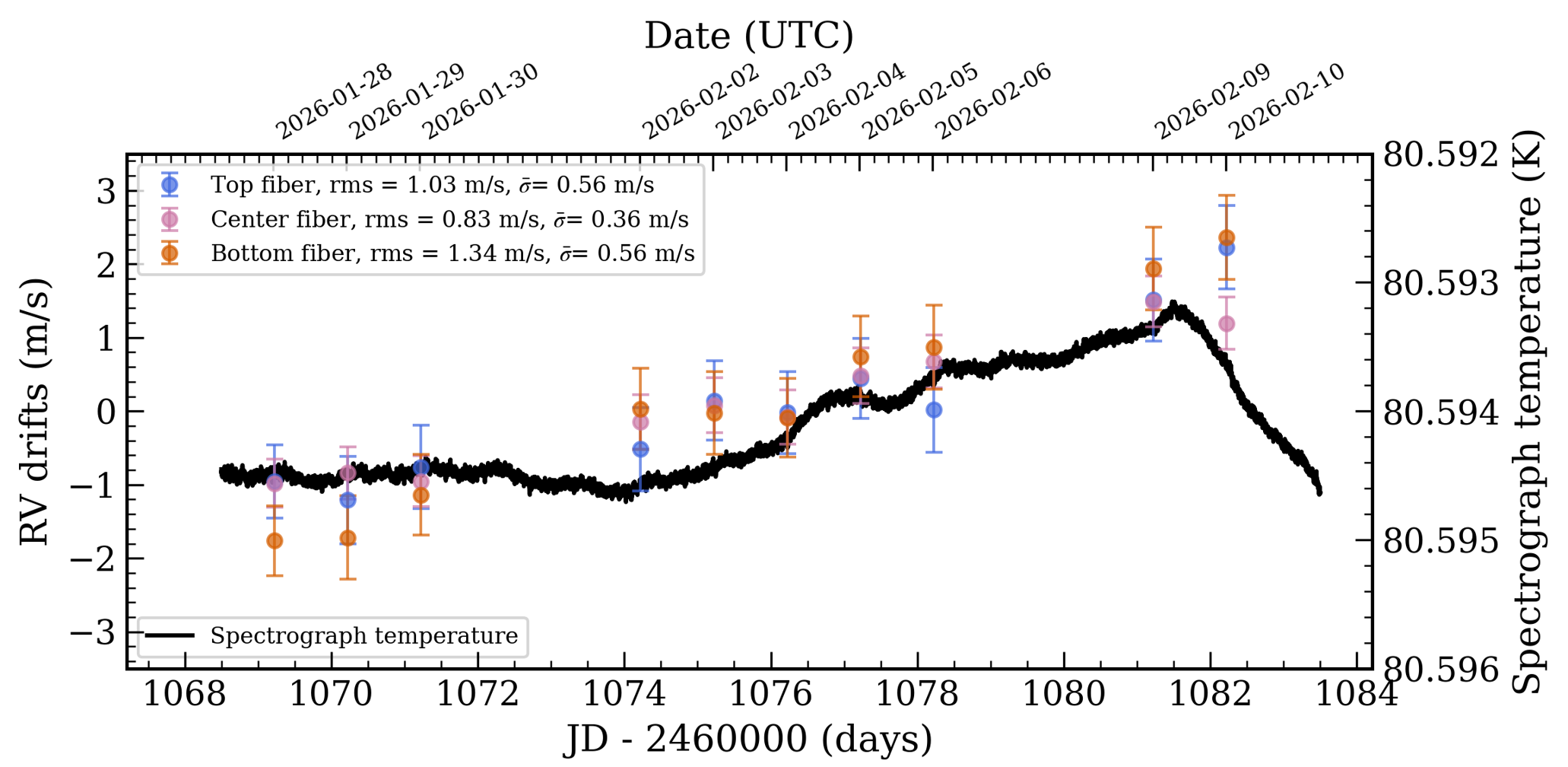}
    \caption{RV drifts of the top, central and bottom fibers shown in blue, pink, and orange, respectively. The spectrograph temperature, measured via a sensor installed on the mount of the echelle grating, is shown in black. Temperature values are centered around the mean, and its axis has been inverted for better visualization of the order of magnitude of the temperature changes and its correlation with the RV drifts.}
    \label{fig:rvstability}
\end{figure}

Figure \ref{fig:rvstability} shows the drift for the three fibers using the line shift method in velocity space. The drift of the three fibers agrees within the errors. The central fiber is the most stable with a rms=0.86 m/s and a mean error of $\Bar{\sigma}$=0.36 m/s.   
The mean error of the top and bottom fibers is $\Bar{\sigma}$=0.56 m/s, slightly larger than the central fiber. This can be explained by the fact that in the laboratory setup, the top and bottom fiber split the flux between them when illuminated (i.e. ~50\%) when compared to flux received by the center fiber (100\%).  This is a limitation of the laboratory configuration and has been resolved by the final fiber switcher which is now in use at the LBT\cite{SPIE2026:calibration}.

\section{CONCLUSIONS}

This paper has presented results from four laboratory characterization activities supporting the development and commissioning of the iLocater spectrograph. The model-based focusing approach using OpticStudio demonstrated improvement in PSF quality relative to the pre-alignment baseline. Detector clocking  correction provided a better alignment of the spectra with respect to the detector pixel grid. The installed LED flat-field system delivers uniform illumination over the full detector. Stability monitoring under lab conditions revealed an instrumental drift below  1 m/s for the central fiber over the course of two weeks, consistent with thermal variations at the level of 1.5 mK. Collectively, these results help validate the instrument's readiness for the next phase of integration at the telescope and support confidence in its science performance at the LBT\cite{SPIE2026:commissioning}.

\acknowledgments % equivalent to \section*{ACKNOWLEDGMENTS}       
 
This material is based upon work supported by the National Science Foundation under Grant No. 1654125, 2108603 and 2408424, the National Aeronautics and Space Administration under Space Act Agreement No. 38232 and Grant No.  80NSSC24K1444 and the Mt. Cuba Astronomical Foundation. JRC acknowledges partial support for the iLocater project from the Wolfe family and Potenziani family.

% References
\bibliography{report} % bibliography data in report.bib

@INPROCEEDINGS{Crass2022_cryo,
       author = {{Crass}, Jonathan and {Sadagopan}, Nandini and {Misch}, Matthew and {Rizika}, Alexa and {Sands}, Brian and {Engstrom}, Matthew and {Crepp}, Justin R. and {Smous}, James and {Chilcote}, Jeffrey and {Fantano}, Louis G. and {VanSickle}, Mike and {Hearty}, Frederick R. and {Nelson}, Matthew J.},
        title = "{The iLocater cryostat and thermal control system: enabling extremely precise radial velocity measurements for diffraction-limited spectrographs}",
    booktitle = {Ground-based and Airborne Instrumentation for Astronomy IX},
         year = 2022,
       editor = {{Evans}, Christopher J. and {Bryant}, Julia J. and {Motohara}, Kentaro},
       series = {Society of Photo-Optical Instrumentation Engineers (SPIE) Conference Series},
       volume = {12184},
        month = aug,
          eid = {121844X},
        pages = {121844X},
          doi = {10.1117/12.2630199},
archivePrefix = {arXiv},
       eprint = {2209.00028},
 primaryClass = {astro-ph.IM},
       adsurl = {https://ui.adsabs.harvard.edu/abs/2022SPIE12184E..4XC}
}

@ARTICLE{Crass2021,
       author = {{Crass}, J. and {Bechter}, A. and {Sands}, B. and {King}, D. and {Ketterer}, R. and {Engstrom}, M. and {Hamper}, R. and {Kopon}, D. and {Smous}, J. and {Crepp}, J.~R. and {Montoya}, M. and {Durney}, O. and {Cavalieri}, D. and {Reynolds}, R. and {Vansickle}, M. and {Onuma}, E. and {Thomes}, J. and {Mullin}, S. and {Shelton}, C. and {Wallace}, K. and {Bechter}, E. and {Vaz}, A. and {Power}, J. and {Rahmer}, G. and {Ertel}, S.},
        title = "{Final design and on-sky testing of the iLocater SX acquisition camera: broad-band single-mode fibre coupling}",
      journal = {mnras},
         year = 2021,
        month = feb,
       volume = {501},
       number = {2},
        pages = {2250-2267},
          doi = {10.1093/mnras/staa3355},
archivePrefix = {arXiv},
       eprint = {2010.13795},
 primaryClass = {astro-ph.IM},
       adsurl = {https://ui.adsabs.harvard.edu/abs/2021MNRAS.501.2250C}
}

@INPROCEEDINGS{Crass2022,
       author = {{Crass}, Jonathan and {Aikens}, David and {Mason}, Joaquin and {King}, David and {Crepp}, Justin R. and {Bechter}, Andrew and {Bechter}, Eric and {Farsad}, Mahsa and {Schwab}, Christian and {VanSickle}, Michael},
        title = "{The final design of the iLocater spectrograph: an optimized architecture for diffraction-limited EPRV instruments}",
    booktitle = {Ground-based and Airborne Instrumentation for Astronomy IX},
         year = 2022,
       editor = {{Evans}, Christopher J. and {Bryant}, Julia J. and {Motohara}, Kentaro},
       series = {Society of Photo-Optical Instrumentation Engineers (SPIE) Conference Series},
       volume = {12184},
        month = aug,
          eid = {121841P},
        pages = {121841P},
          doi = {10.1117/12.2630228},
archivePrefix = {arXiv},
       eprint = {2209.00009},
 primaryClass = {astro-ph.IM},
       adsurl = {https://ui.adsabs.harvard.edu/abs/2022SPIE12184E..1PC}
}

@INPROCEEDINGS{Crepp2016,
       author = {{Crepp}, Justin R. and {Crass}, Jonathan and {King}, David and {Bechter}, Andrew and {Bechter}, Eric and {Ketterer}, Ryan and {Reynolds}, Robert and {Hinz}, Philip and {Kopon}, Derek and {Cavalieri}, David and {Fantano}, Louis and {Koca}, Corina and {Onuma}, Eleanya and {Stapelfeldt}, Karl and {Thomes}, Joseph and {Wall}, Sheila and {Macenka}, Steven and {McGuire}, James and {Korniski}, Ronald and {Zugby}, Leonard and {Eisner}, Joshua and {Gaudi}, B.~S. and {Hearty}, Fred and {Kratter}, Kaitlin and {Kuchner}, Marc and {Micela}, Giusi and {Nelson}, Matthew and {Pagano}, Isabella and {Quirrenbach}, Andreas and {Schwab}, Christian and {Skrutskie}, Michael and {Sozzetti}, Alessandro and {Woodward}, Charles and {Zhao}, Bo},
        title = "{iLocater: a diffraction-limited Doppler spectrometer for the Large Binocular Telescope}",
    booktitle = {Ground-based and Airborne Instrumentation for Astronomy VI},
         year = 2016,
       editor = {{Evans}, Christopher J. and {Simard}, Luc and {Takami}, Hideki},
       series = {Society of Photo-Optical Instrumentation Engineers (SPIE) Conference Series},
       volume = {9908},
        month = aug,
          eid = {990819},
        pages = {990819},
          doi = {10.1117/12.2233135},
archivePrefix = {arXiv},
       eprint = {1609.04412},
 primaryClass = {astro-ph.IM},
       adsurl = {https://ui.adsabs.harvard.edu/abs/2016SPIE.9908E..19C}
}

@article{zospy24, 
author = {van Vught, Luc and Haasjes, Corné and Beenakker, Jan-Willem M.}, 
title = {ZOSPy: optical ray tracing in Python through OpticStudio},
doi = {10.21105/joss.05756}, 
url = {https://doi.org/10.21105/joss.05756}, 
year = {2024}, 
publisher = {The Open Journal}, 
volume = {9}, 
number = {96}, 
pages = {5756},
journal = {Journal of Open Source Software} 
}

@inproceedings{HxRGproc_Ninan,
author = {J. P. Ninan and Chad F. Bender and Suvrath Mahadevan and Eric B. Ford and Andrew J. Monson and Kyle F. Kaplan and Ryan C. Terrien and Arpita Roy and Paul M. Robertson and Shubham Kanodia and Gudmundur K. Stefansson},
title = {{The Habitable-Zone Planet Finder: improved flux image generation algorithms for H2RG up-the-ramp data}},
volume = {10709},
booktitle = {High Energy, Optical, and Infrared Detectors for Astronomy VIII},
editor = {Andrew D. Holland and James Beletic},
organization = {International Society for Optics and Photonics},
publisher = {SPIE},
pages = {107092U},
year = {2018},
doi = {10.1117/12.2312787},
URL = {https://doi.org/10.1117/12.2312787}
}

@article{HxRGproc_Rauscher,
doi = {10.1086/684082},
url = {https://doi.org/10.1086/684082},
year = {2015},
month = {nov},
publisher = {The Astronomical Society of the Pacific},
volume = {127},
number = {957},
pages = {1144},
author = {Rauscher, Bernard J.},
title = {Teledyne H1RG, H2RG, and H4RG Noise Generator},
journal = {Publications of the Astronomical Society of the Pacific}
}

@ARTICLE{Horne1986,
       author = {{Horne}, K.},
        title = "{An optimal extraction algorithm for CCD spectroscopy.}",
      journal = {PASP},
         year = 1986,
        month = jun,
       volume = {98},
        pages = {609-617},
          doi = {10.1086/131801},
       adsurl = {https://ui.adsabs.harvard.edu/abs/1986PASP...98..609H}
}

@ARTICLE{redman2011,
       author = {{Redman}, Stephen L. and {Lawler}, James E. and {Nave}, Gillian and {Ramsey}, Lawrence W. and {Mahadevan}, Suvrath},
        title = "{The Infrared Spectrum of Uranium Hollow Cathode Lamps from 850 nm to 4000 nm: Wavenumbers and Line Identifications from Fourier Transform Spectra}",
      journal = {apjs},
         year = 2011,
        month = aug,
       volume = {195},
       number = {2},
          eid = {24},
        pages = {24},
          doi = {10.1088/0067-0049/195/2/24},
archivePrefix = {arXiv},
       eprint = {1107.4091},
 primaryClass = {astro-ph.IM},
       adsurl = {https://ui.adsabs.harvard.edu/abs/2011ApJS..195...24R}
}

@INPROCEEDINGS{SPIE2026:calibration,
author = {{Senthil Nathan}, Sai Vidyud and {Crass}, Jonathan and {Brady}, Julia and {Derwent}, Mark and {Tala Pinto}, Marcelo Said and {Sands}, Brian and {Datkuliak}, Jackson and {Shover}, Jonathan and {Schwab}, Christian and {Stürmer}, Julian and {Letchev}, Stanimir O. and {Pember}, Jacob and {Spiegel}, Jayde and {Duell}, Erin and {Pappalardo}, Daniel and {Johnson}, Marshall C. and {Engelman}, Michael and {Castro}, Matheus de Jesus and {Crepp}, Justin R. and {Kitzler}, Ondrej and {Legero}, Thomas and {Lesley}, Xavier and {Pogge}, Richard and {Zielinski-Nicolson}, Dane},
title = "{The Calibration System of the iLocater Spectrograph}",
booktitle = {Ground-based and Airborne Instrumentation for Astronomy XI},
year = 2026,
editor = {{Vernet}, Jo\:el and {Bryant}, Julia J. and {Motohara}, Kentaro},
series = {Society of Photo-Optical Instrumentation Engineers (SPIE) Conference Series},
volume = {14149},
month = aug,
eid = {14149-462},
pages = {},
doi = {},
}

@ARTICLE{Crepp2014,
       author = {{Crepp}, Justin R.},
        title = "{Improving planet-finding spectrometers}",
      journal = {Science},
         year = 2014,
        month = nov,
       volume = {346},
       number = {6211},
        pages = {809-810},
          doi = {10.1126/science.1262071},
archivePrefix = {arXiv},
       eprint = {1412.2992},
 primaryClass = {astro-ph.IM},
       adsurl = {https://ui.adsabs.harvard.edu/abs/2014Sci...346..809C}
}

@INPROCEEDINGS{Esposito2011,
       author = {{Esposito}, S. and {Riccardi}, A. and {Pinna}, E. and {Puglisi}, A. and {Quir{\'o}s-Pacheco}, F. and {Arcidiacono}, C. and {Xompero}, M. and {Briguglio}, R. and {Agapito}, G. and {Busoni}, L. and {Fini}, L. and {Argomedo}, J. and {Gherardi}, A. and {Brusa}, G. and {Miller}, D. and {Guerra}, J.~C. and {Stefanini}, P. and {Salinari}, P.},
        title = "{Large Binocular Telescope Adaptive Optics System: new achievements and perspectives in adaptive optics}",
    booktitle = {Astronomical Adaptive Optics Systems and Applications IV},
         year = 2011,
       editor = {{Tyson}, Robert K. and {Hart}, Michael},
       series = {Society of Photo-Optical Instrumentation Engineers (SPIE) Conference Series},
       volume = {8149},
        month = oct,
          eid = {814902},
        pages = {814902},
          doi = {10.1117/12.898641},
archivePrefix = {arXiv},
       eprint = {1203.2761},
 primaryClass = {astro-ph.IM},
       adsurl = {https://ui.adsabs.harvard.edu/abs/2011SPIE.8149E..02E}
}

@INPROCEEDINGS{SPIE2026:commissioning,
author = {{Johnson}, Marshall C. and {Tala Pinto}, Marcelo and {Crass}, Jonathan and {Sands}, Brian and {Datkuliak}, Jackson and {Senthil Nathan}, Sai Vidyud {Crepp}, Justin R. and {Engelman}, Michael and {Pappalardo}, Daniel and {Brady}, Julia E. and {Derwent}, Mark and {Lesley}, Xavier and {Bechter}, Andrew and {Bechter}, Eric and {Schwab}, Christian and {Aikens}, David and {Bender}, Chad F. and {Brandon}, Chris and {Brooks}, Cynthia and {Castro}, Matheus J. and {Chilcote}, Jeffrey and {Conrad}, Al and {Delo}, Gregory and {Duell}, Erin and {Engstrom}, Matthew and {Ertel}, Steve and {Fantano}, Louis and {Feggans}, John Keith and {Flores}, Hali and {Gaudi}, B. Scott and {King}, David and {Kitzler}, Ondrej and {Kratter}, Kaitlin and {Kruk}, Jeffrey and {Legero}, Thomas and  {Letchev}, Stanimir O. and {Mason}, Jerry and {Mason}, Joaquin and {Misch}, Matthew and {Pember}, Jacob and {Pogge}, Richard and {Quirrenbach}, Andreas and {Schlieder}, Joshua and {Waczynski}, Augustyn and {Shields}, Joseph and {Kim}, Daewook and {Smithwright}, Mark and {Renshaw}, Allie and {Richie}, Austin and {Rizika}, Alexa and {Sadagopan}, Nandini and {Shover}, Jonathan and {Smous}, James and {Spiegel}, Jayde and {Stürmer}, Julian and {Sura}, Sahil and {Switzer}, Robert and {Thomes}, Joe and {Wang}, Ji and {Weiss}, Lauren and {Wilson}, Daniel and {Zielinski-Nicolson}, Dane},
title = "{Commissioning and on-sky performance verification of iLocater}",
booktitle = {Ground-based and Airborne Instrumentation for Astronomy XI},
year = 2026,
editor = {{Vernet}, Jo\:el and {Bryant}, Julia J. and {Motohara}, Kentaro},
series = {Society of Photo-Optical Instrumentation Engineers (SPIE) Conference Series},
volume = {14149},
month = aug,
eid = {14149-195},
pages = {},
doi = {},
}
\bibliographystyle{spiebib} % makes bibtex use spiebib.bst

\end{document}